\documentclass{elsart}

\usepackage[english]{babel}
\usepackage[latin2]{inputenc}
\usepackage[T1]{fontenc}

\usepackage{ae,aecompl}
\usepackage{epsfig}
\usepackage{graphics}
\usepackage{graphicx}
\usepackage{amsmath}
\usepackage{amssymb}
\usepackage{pifont}

\begin{document}

\begin{frontmatter}
\title{Multifractal model of asset returns with leverage effect}
\author[bme]{Z. Eisler},
\ead{eisler@maxwell.phy.bme.hu}
\author[bme,lce]{J. Kertész}
\address[bme]{Department of Theoretical Physics, Budapest University of Technology and Economics, Budafoki út 8, H-1111 Budapest, Hungary}
\address[lce]{Laboratory of Computational Engineering, Helsinki University of Technology, P.O.Box 9400, FIN-02015 HUT, Finland}
\begin{abstract}
Multifractal processes are a relatively new tool of stock market analysis. Their power lies in the ability to take multiple orders of autocorrelations into account explicitly. In the first part of the paper we discuss the framework of the Lux model and refine the underlying phenomenological picture. We also give a procedure of fitting all parameters to empirical data. We present a new approach to account for the effective length of power-law memory in volatility. The second part of the paper deals with the consequences of asymmetry in returns. We incorporate two related stylized facts, skewness and leverage autocorrelations into the model. Then from Monte Carlo measurements we show, that this asymmetry significantly increases the mean squared error of volatility forecasts. Based on a filtering method we give evidence on similar behavior in empirical data.
\end{abstract}
\end{frontmatter}

\section{Introduction}
The past decades of quantitative analysis of stock market data have
produced several fruitful concepts and predictive methods. After the
groundbreaking works of Bachelier \cite{bachelier} and with the early
discovery of the broad distribution of asset returns \cite{mandelbrot.certain}, the application of additional insights from physics have become possible \cite{stanley.universality2,stanley.contribute,ivory}. Similarly to the case of fully developed turbulence \cite{vicsek.turb}, multifractal models \cite{mandelbrot.mmar,lux.model,mrw.pre,schmitt} catch important aspects of price patterns, that the traditional ARCH-GARCH \cite{arch,garch,figarch} framework cannot \cite{mandelbrot.mmar}. Scaling methods \cite{vicsek.book,dfa.intro,dfa.kantelhardt} can be used to quantify the autocorrelation properties of these patterns, and the results can be taken into account explicitly. Based on a discrete time stochastic volatility framework, one can construct powerful and flexible random processes, which have great applicability to forecasting.

Present paper deals with the applications of the model introduced in
\cite{lux.model}. In the following we will refer to this as the
\emph{Lux model}, because we follow exactly the framework developed by
Thomas Lux. We must note, that based on the paper by Mandelbrot,
Calvet and Fisher \cite{mandelbrot.mmar}, a similar model was proposed
by Calvet and Fisher \cite{cf.forecasting}. Section \ref{sec:stylized}
facilitates notations and describes the main important stylized facts
of stock prices. Section \ref{sec:logret} introduces discrete time
stochastic volatility models and the Lux model of stock returns. We
give some additional insights into the underlying phenomenology,
rooted in empirical findings \cite{stanley.trading2}. Section
\ref{sec:lux} gives one method of parameter approximation and
discusses limitations of the Lux model. Section \ref{sec:third} deals
with the extension of the model to account for skewness and leverage
autocorrelations
\cite{leverage1,mrw.skewness,leverage3,leverage4}. Based on a
corresponding filtering of the data and Monte Carlo measurements we
evaluate the effect of asymmetry on volatility forecasting.

The calculations were compared to an analysis of real financial data.
We analyzed the Deutsche Aktienindex (\emph{DAX}) in the period
Jan. 1999 to Dec. 2001. The data includes transactions before opening
and after closing. The sampling interval is $15$ seconds. The series
consists of $1829545$ data points. We also analyzed data of the $200$
most liquid (traded in most minutes over the interval) stocks at New
York Stock Exchange (\emph{NYSE200}). The data is based on the TAQ
database \cite{taq}, and it includes the period Jan. 2000 -
Sep. 2002. The sampling interval is $1$ minute, with a total of
$266966$ data points. 
\section{Empirical properties of stock returns: stylized facts}
\label{sec:stylized}

Over the last century of stock market research and especially in the last few decades, a number of universal features 
regarding stock market price dynamics have become apparent \cite{stanley.book,bouchaud.book,kertesz.book,cont.stylized}. These so-called \emph{stylized facts} are (at least qualitatively) independent of the market, the asset traded\footnote{There are some notable differences between stocks, commodities and foreign exchange data, however these are beyond the scope of our paper. In the following we will limit ourselves to the discussion of stock price properties, although most observations hold for various markets. One remarkable difference is the lack of long-term up-down asymmetry in foreign exchange price variations, a consequence of a greater symmetry than usual stock versus currency valuation \cite{cont.stylized}.} and even the time period, although their quantitative value might vary due to inevitable instationarities.

Let $Y(t)$ denote the price of a financial asset at time $t$. We define \emph{returns} as 
\begin{equation}
r_{\Delta t}(t)=Y(t+\Delta t)-Y(t).
\end{equation}
The logarithm of the price will be denoted as $p(t)=\log (Y(t))$. The \emph{logarithmic returns} over a time horizon $\Delta t$ are then defined as 
\begin{equation}
x_{\Delta t}(t)=p(t+\Delta t)-p(t)\equiv \log\left(\frac{Y(t+\Delta t)}{Y(t)}\right).
\end{equation}
The first major group of stylized facts of our concern is related to the probability distribution of log-returns $x_{\Delta t}$ (see also Figure \ref{fig:dax-lux-dist}). Classical stylized facts for short (minute-scale) $\Delta t$ are that the center of the distribution (density function) behaves like a Lévy distribution ($f(r)\propto r^{-1}$), and that the tails are truncated according to $f(r)\propto r^{-\alpha}$ with $\alpha \in [3, 6]$ \cite{stanley.trading2,stanley.statistical,stanley.scaling,stanley.individual}. This is the so-called ``inverse cubic law''. The tail exponent for DAX was found to be $\alpha \approx 3.79$ (see also Figure \ref{fig:dax-lux-dist}). This decay is much slower, than expected from a Brownian approximation: many-$\sigma$ events (large price changes) are not uncommon. These heavy tails can be characterized by the \emph{kurtosis}:
\begin{equation}
\kappa_4 = \frac{\left\langle \left(x-\left\langle x \right\rangle\right)^4\right\rangle}{\left\langle x^2\right\rangle^2}.
\end{equation}
Gaussian distribution has a kurtosis of $3$, a greater value shows an excess weight of large events. At values of $\alpha \leq 5$ kurtosis does not exist, and the measured values originate from a finite sample effect. The estimate of $\kappa_4$ diverges as the length of the measurement period goes to infinity \cite{stanley.trading2}.

The distinct scaling behavior of the tails is very useful for applications, because that regime corresponds to large events \cite{kertesz.drops}, the main sources of financial risk. It is notable, that the power-law regime of \emph{DAX} is observed to be $20\sigma$-s wide, that is much greater than usual tails of the Gaussian distribution. We have to note on the immense predictive power of this observation. Over the whole three years of observation the number of $6\sigma +$ events was approximately $200$. For a shorter time series given, the data is even more scarce, and we cannot give a statistically correct approximation of the large event distribution, unless we accept this kind of extrapolation.

Classical studies also imply, that the distribution has a negative \emph{skewness}, defined as: 
\begin{equation}
\kappa_3 = \frac{\left\langle \left(x-\left\langle x \right\rangle\right)^3\right\rangle}{\left\langle x^2\right\rangle^{3/2}} < 0.
\end{equation}
The presence of negative skewness is called \emph{gain-loss asymmetry}
\cite{cont.stylized}, and it is attributed to psychological
factors. More recent analyses show \cite{rally} that the skewness is not stationary, but it characterizes the state of the market: it is negative when a downward tendency of prices is present, but positive on rally days.

All the above apply to short time scales, nevertheless the distribution is not stable. By increasing the $\Delta t$ horizon, slowly, with a characteristic time horizon of $\Delta t \sim 10-100$ days, it converges to a Gaussian. This is called \emph{aggregational Gaussianity} \cite{cont.stylized,stanley.individual,kullmann1,kullmann2}. Consequently, also: $\kappa_3 \rightarrow 0$, $\kappa_4 \rightarrow 3$. This convergence is not unexpected from the central limit theorem, as $\left\langle x_{\Delta t}^2(t)\right\rangle$ exists \cite{cont.stylized}.

A second major group of stylized facts is related to the autocorrelation properties of the time series. These also display universal features (also with some differences between stocks, commodities, and foreign exchange rates \cite{cont.stylized,stanley.commodities}). These properties also appear to be very robust for different time periods investigated. 
The returns are said to have a \emph{short memory}. This means, that their \emph{linear} autocorrelation decays exponentially: 
\begin{equation}
C_r(\tau)=\left\langle r_{\Delta t}(t+\tau)r_{\Delta t}(t)\right\rangle\propto \exp\left(-\frac{t}{\tau}\right),
\end{equation}
where $\tau$ is of the order of $10$ minutes \cite{stanley.statistical,stanley.scaling}. This, however, by far does not mean, that they are independent. Non-linear functions of the returns do exhibit long-memory. Regarding our later arguments, we found useful to measure the autocorrelations of squared logarithmic returns. This can be given as:
\begin{equation}
\label{eq:x2x2ac}
C_{x^2}(\tau)=\left\langle x_{\Delta t}^2(t+\tau)x_{\Delta t}^2(t)\right\rangle .
\end{equation}
Measurements on \emph{NYSE200} show, that for $\Delta t < 1$ day, $C_{x^2}(\tau ) \propto \tau ^{-\beta}$, with $\beta \approx 0.307$. Apparently $C_{x^2}(\tau \rightarrow \infty) = \sigma^4$, where $ \sigma^2 = \left\langle x_{\Delta t}(t)^2\right\rangle$. Note that the previously mentioned aggregational Gaussianity is a stronger statement, than just the disappearance of correlations. Nevertheless, this can be one good indicator of such a crossover. Other signs could be the breakdown of multifractal scaling or the disappearance of excess kurtosis. These have a similar time scale, but there is no theoretical necessity for all to occur at the very same $\Delta t$. In fact there are strong indications, that the time of the crossover point depends on the criterion we choose \cite{cont.stylized,kullmann2}. 

There is a strong connection with two more stylized facts:
\emph{volatility clustering} and \emph{volatility mean reversion}. By
volatility we will refer to the instantaneous standard deviation of
log-returns. There is assumed to be a "normal", mean level of
volatility, that characterizes $x$ over a long time interval. If a
difference from this average behavior emerges, the alternative size
log-returns tend to be clustered together, corresponding to greater
volatility for a time period. Then eventually this excess volatility
decays to the mean level. This is the process which can be quantified 
by $C_{x^2}(\tau )$.

One can define generalized $q$-th order autocorrelation functions as
in \eqref{eq:tqdef}, and it is widely recognized, that the related
\emph{scaling exponents} $\tau (q)$ are a non-linear\footnote{With
  this definition there is a linear relation (also referred to as \emph{monofractality}) for Brownian motion: $\tau_{Brownian} (q)=q/2-1$. This can be taken as a good reference for measurements.} function of $q$. This is called the \emph{multifractal} nature of logarithmic (and also linear) returns \cite{schmitt,cont.stylized,stanley.commodities,mandelbrot.exchange}. The power-law scaling law holds for very long (months to years) time lags \cite{ausloos}.
\begin{equation}
c_q(\Delta t) \equiv \left \langle \vert x_{\Delta t}(t)\vert^q \right \rangle \propto \Delta t^{\tau(q)+1}
\label{eq:tqdef}
\end{equation}
The last stylized fact we would like to mention is the \emph{leverage effect}\footnote{Note that this seriously differs from what traders refer to as leverage.}, that is closely related to gain-loss asymmetry. As reported by several works, for positive time lags the volatility-return correlation function is asymmetric and decays with time \cite{leverage1,leverage3,leverage4,cont.stylized,bouchaud.leverage}:
\begin{equation}
\label{eq:leverage}
C_L(\tau ) = \frac{\left\langle x_{\Delta t}(t) x_{\Delta t}^2 (t+\tau)\right\rangle}{\left\langle x_{\Delta_t}^2(t) \right\rangle^2} \approx -\Theta(\tau )K^*(\tau ).
\end{equation}
$\Theta (\Delta t)$ is the Heaviside function. The decay is usually considered exponential. $K^*(t)= A\exp\left(-\frac{t}{\tau_0}\right )$, where \cite{bouchaud.leverage} reports $A\approx 1.9$ and $\tau \approx 69$ days for American stocks. On the other hand these results are usually very noisy, and in the pertinent literature we find no indisputable argument for the exponentiality. There are various, qualitatively similar formulas originating from different models \cite{leverage4}. In the case of DAX we find that data are just as well fitted by a power-law, $K^*(t) = B t^{-\alpha^*}$. 

\section{Modeling logarithmic returns}
\label{sec:logret}
\subsection{Stochastic volatility models}
\label{sec:stochvol}
Logarithmic returns are considered in this framework to originate from the product of two time series: 
\begin{equation}
x_{\Delta t}(t)=A_{\Delta t}(t)\sigma S_{\Delta t}(t).
\label{eq:volat}
\end{equation}
$\sigma$ is the standard deviation of logarithmic returns, in our notations $A(t)$ and $S(t)$ are normalized to unit mean square. $S(t)$ is a sign factor, usually assumed to be uncorrelated. $A(t)$ is an amplitude factor, usually it is defined to be positive, and it is long-time correlated. This latter process models basically the \emph{volatility}.  It is the source of volatility clustering, mean reversion and time correlations apparent in returns (see Section \ref{sec:stylized}). As $S(t)$ practically determines the sign of $x$ only, volatility can be approximated as 
\begin{equation}
\sigma A_{\Delta t}(t)=\left|x_{\Delta t}(t)\right|.
\label{eq:stochvol}
\end{equation}
The sign process $S(t)$ is mostly modeled by Gaussian white noise, or simply a random $+1$/$-1$ value.

Recent studies have unveiled a strong relationship between trading activity and the volatility process. Plerou et al. \cite{stanley.trading2} imply, that on any time scale $\Delta t$, where the ${\mathfrak N}_{\Delta t}(t)$ number of trades of the asset in the periods $[t,t+\Delta t]$ is much greater, than $1$ (for liquid stocks they find this limit applies from as short as $\Delta t = 15$ minutes), the log-returns of the asset can be well approximated as
\begin{equation}
x_{\Delta t}(t) = \sqrt{{\mathfrak N}_{\Delta t}(t)}W_{\Delta t}(t)N(0,1).
\label{eq:trade}
\end{equation}
${\mathfrak N}_{\Delta t}(t)$ is found to be strongly correlated, and to have power-law tails. The term $W_{\Delta t}(t)$ is the realized volatility, which is only weakly correlated, and also has power-law tails. 
\subsection{Multifractal modeling}

One can define a broad class of models of logarithmic returns through
\eqref{eq:stochvol}. The aim is to find an appropriate $A(t)$
stochastic process, that accounts for as many stylized facts as
possible. We consider here models which focus on the multifractal
property \eqref{eq:tqdef}, because this is a general approach to take
multiple orders of autocorrelations into account. For practical
applications like volatility forecasting these play a key role. In the
literature, there are three relevant multifractal models currently
available. The first proposition to use a multifractal process as
stochastic volatility was the Multifractal Model of Asset Return
(MMAR) by Mandelbrot, Calvet and Fisher given in a series of papers
\cite{mandelbrot.mmar,mandelbrot.exchange,mandelbrot.large}. A
different process, the Multifractal Random Walk (MRW) was developed in
\cite{mrw.pre,mrw.epjb,mrw.physa,endoexo}. Here we discuss in detail a
modified version of the MMAR model which we call Lux model  \cite{lux.model}.

\subsection{The Lux model}
In both MMAR and the Lux model the logarithmic price is assumed to follow
\begin{equation}
p(t) = p(0) + B(\theta (t)),
\label{eq:mmar}
\end{equation}
where $B(t)$ is an Integral Brownian motion with standard deviation
$\sigma$ and $\theta (t)$ is an increasing function of $t$, and it plays the role of time-deformation, which is well-established in econometric modeling \cite{mandelbrot.certain,mandelbrot.mmar,timedef1,timedef2}.

\begin{figure}[!t]
\hbox{\psfig{file=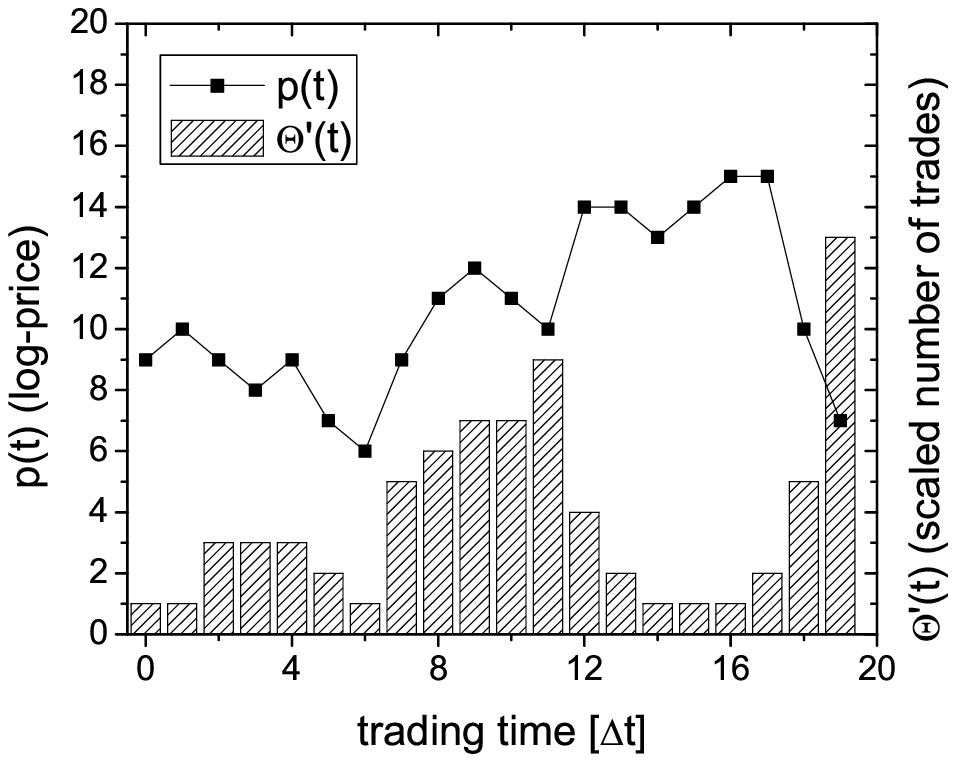,height=170pt}\psfig{file=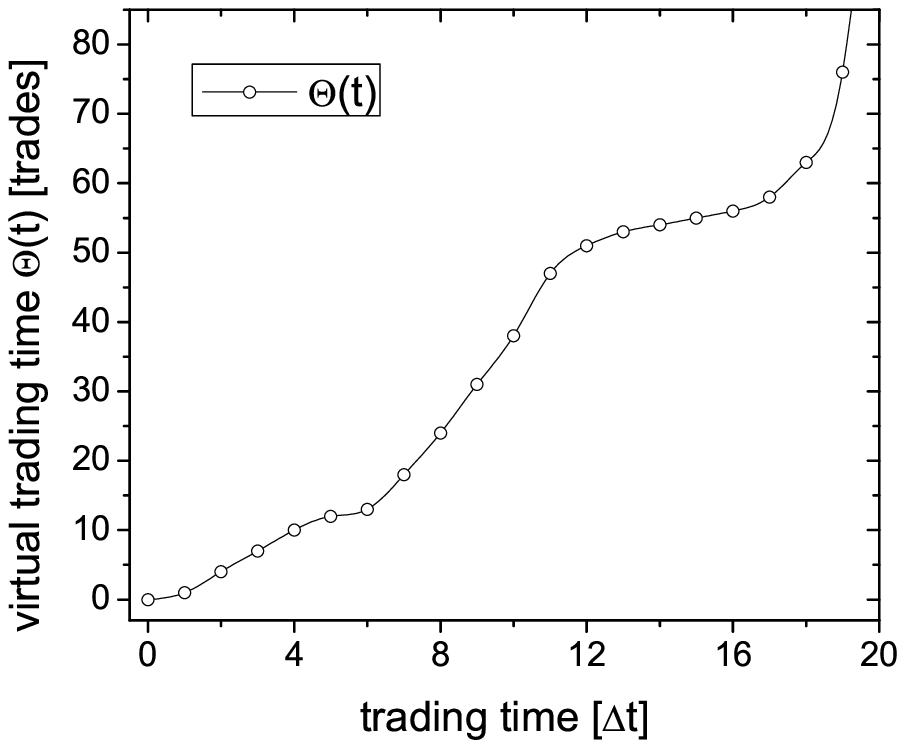,height=170pt}}
\caption{A visual explanation of the common idea of MMAR and the Lux model. The actual values are only to be taken as illustration. Left: $p(t)$ logarithmic price of the asset and the $\theta' (t)$ increments of trading time. One can see, that the absolute value of log-returns is usually larger in periods, where $\theta'(t)$ is larger. Right: The corresponding $\theta (t)$ flow of virtual trading time as a function of real time.}
\label{fig:lux-explain}
\end{figure}

Let $\theta'(t)$ denote the increments $\theta'(t)=\theta
(t)-\theta(t-\Delta t)$. The function $\theta (t)$ can be thought of
as a virtual trading time, that increases with the real time $t$. In
different periods the passing of $\theta (t)$ can be slower or faster
(a visual explanation for the process can be seen in Figure
\ref{fig:lux-explain}). When measured in this virtual time, the
logarithmic price is assumed to follow a simple Integral Brownian
Motion, and thereby slower or faster trading (smaller and larger
values of $\theta'(t)$ respectively) correspond to smaller and greater
volatility from the real time point of view. The finite lengths of
these regimes result in volatility clustering. 

To gain deeper insight into the meaning of time-deformation we have to reformulate \eqref{eq:mmar}. We know, that in discrete time $B(t)$ is the sum of independent Gaussian increments $\sigma S(t) = N(0, \sigma)$. Discretization of time can be interpreted as having a finite sampling interval of financial data (minutes or days). Using the stability of the Gaussian distribution one can give a somewhat generalized formalization:
\begin{equation}
B(\theta (t)) = \sigma N(0,\sqrt{\theta (t)/\Delta t}).
\end{equation}
With the above notations we get\footnote{One can see, that we can choose $\Delta t = 1$, without any loss of generality. It is very intuitive, that on a different $\Delta t$ time scale the process remains the same, only the typical size of price changes varies. This is in harmony with the notion of scale-invariance.}:
\begin{equation}
B(\theta(t)) = \sum_{\tau = \Delta t,2\Delta t\dots, t}\sqrt{\theta'(\tau )}\frac{\sigma}{\sqrt{\Delta t}}
N\left (0,1\right ).
\end{equation}
Equivalently for returns:
\begin{equation}
x_{\Delta t}(t) = \sqrt{\theta'(t)}\frac{\sigma}{\sqrt{\Delta t}} N\left (0,1\right ),
\label{eq:trade.mmar}
\end{equation}
where the first term corresponds to the amplitude $A_{\Delta t}(t)$ introduced in \eqref{eq:volat}.

One can deduce a formal analogy between \eqref{eq:trade.mmar} and \eqref{eq:trade}. The value of the $\theta'(t)$ increments of virtual trading time can hence be related to measurable quantities:
\begin{equation}
\sqrt{\theta'(t)}\frac{\sigma}{\sqrt{\Delta t}} = \sqrt{{\mathfrak N}_{\Delta t}(t)}W_{\Delta t}(t).
\label{eq:trade2}
\end{equation}
This shows, that $\theta '(t)$ does not represent purely a time deformation, such as virtual trading time would equal the number of trades. It also contains the realized volatility process $W_{\Delta t}$.

We can think of ${\theta'(t)}$ both as the square volatility on an equidistant time scale, and as a quantity proportional to number of trades in the same period. We discussed in Section \ref{sec:stylized}, that volatility has a multifractal property, but analogously ${{\mathfrak N}_{\Delta t}(t)}$ also appears to lack a typical time scale for highly capitalized and frequently traded stocks \cite{mantegna.trading}.

To complete our definition we have to give the random process we wish to use for modeling $\theta (t)$. It is more convenient to give the increments instead:
\begin{equation}
\theta' (t) = 2^k \prod_{i=1}^k m^{(i)}_{t} .
\end{equation}
$k$ can be any integer number, while all $m^{(i)}_t$ are taken from a log-normal distribution. The trick that facilitates multifractal behavior is that $m_t^{(i)}$ are defined to have strong time autocorrelation:
\begin{equation}
m_{t+1}^{(i)} = \bigg\{\begin{array}{ll} \exp(N(-\lambda\ln 2,2(\lambda - 1)\ln 2))) \\ m_t^{(i)}  \end{array},
\end{equation}
where the choice is made as follows. We go from $i=1\dots k$, and choose the top option with probability $2^{-k+i}$, or if for any previous $i$ we had already chosen the top option. We choose the bottom option otherwise. The moments of the distribution:
\begin{equation}
\left\langle \left(m_t^{(i)}\right)^{q/2}\right\rangle = \exp(-(q/2)\lambda\ln 2) \exp((q/2)^2(\lambda - 1)\ln 2) = 2^{-(\tau_x(q)+1)}.
\label{eq:lux-moments}
\end{equation}
It is easy to show, that the autocorrelation of any $m^{(i)}$ is exponential with characteristic time steps $2^{k-i}$. On the other hand as one takes the $k\rightarrow \infty$ limit, the autocorrelation of $\theta'(t)$ becomes decaying as a power-law. Also for any finite $k < \infty$ the process has the multifractal property \eqref{eq:tqdef} if $\Delta t < 2^k$, with
\begin{equation}
\tau_x (q) = \frac{q}{2}-1-\left(\frac{q^2}{4}-\frac{q}{2}\right)\left(\lambda - 1\right ),
\label{eq:lux-tq}
\end{equation}
the same scaling exponents as for the moments.
\subsection{Issues of causality}
\label{sec:causality}
Now we turn to the question of applicability, as it is very important to note on a major weakness of multifractal models. The MMAR is said to be \emph{not causal}, because the whole return time series is generated at the same time, as a binomial cascade \cite{mandelbrot.mmar}. Hence it cannot be used for volatility forecasting. There is no way to treat empirical data as the "past" values of such a cascade and generate a mean future behavior. The Lux model (and MRW) has no such theoretical limitation, these processes can be generated "causally". This means, that if we know the present state of the model, we can generate the distribution of future conditional volatility step by step. The problem here is how to reconstruct the initial conditions of the model from empirical data. For GARCH-type models \cite{garch} the present state of the market is fully characterized by the past return time series and this data is readily available. The application of the Lux model however would demand the value of the initial $m_{t=0}^{(i=1\dots k)}$ multipliers, not only their products, which are proportional to returns. Multipliers are not directly observable and there is no maximum likelihood estimator available at present. One partial success has been achieved by Calvet and Fisher \cite{cf.regime} recently, but that corresponds to a simplified case. A future advantage of the Lux model can be that the number of these state variables is small ($k$), while for MRW this number is untreatably large (and infinite in the ideal case).

\section{The Lux model revisited}
\label{sec:lux}
The aim of this chapter is to give an outline of the estimation of the parameters of the Lux model. We propose a new method to take the effective length of the power-law memory of log-returns into account explicitly. Then we note on some incosistencies with real data.

\subsection{Parameter approximation}

\begin{figure}[!t]
\hbox{\psfig{file=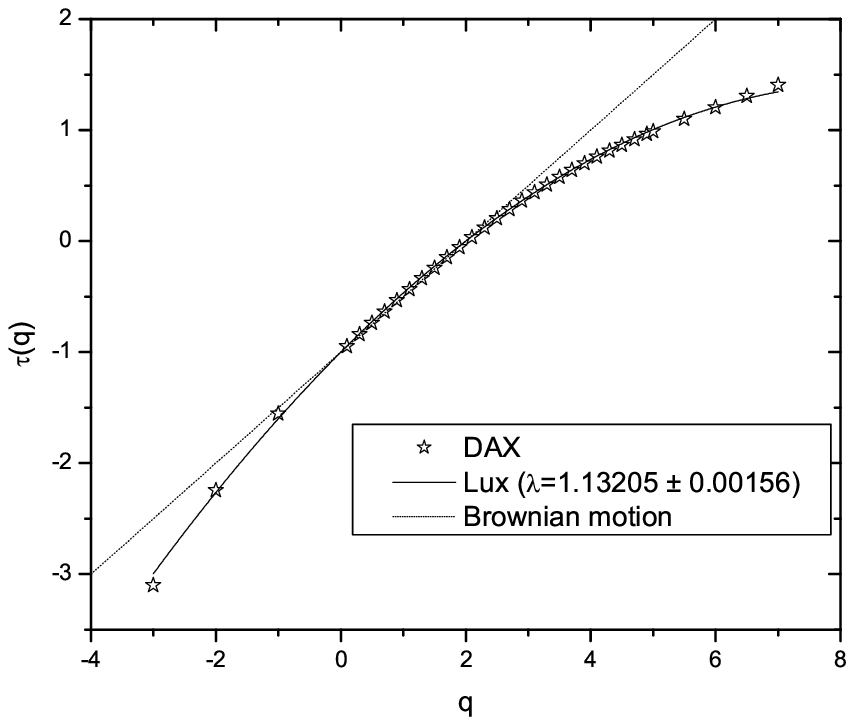,width=250pt}\vbox{\psfig{file=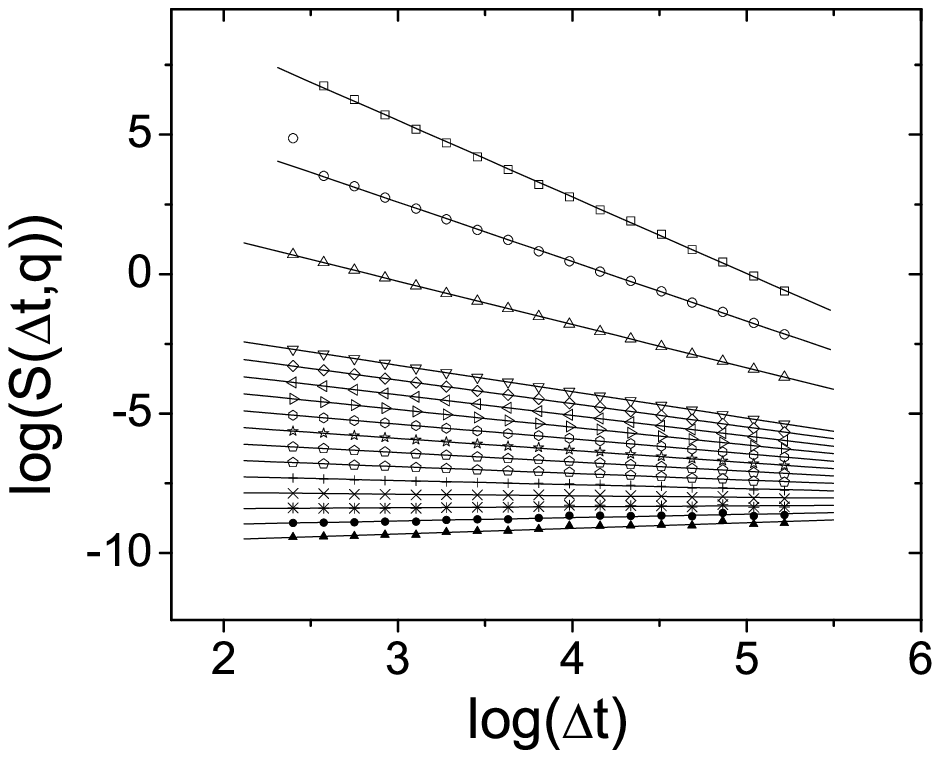,width=125pt}\\\psfig{file=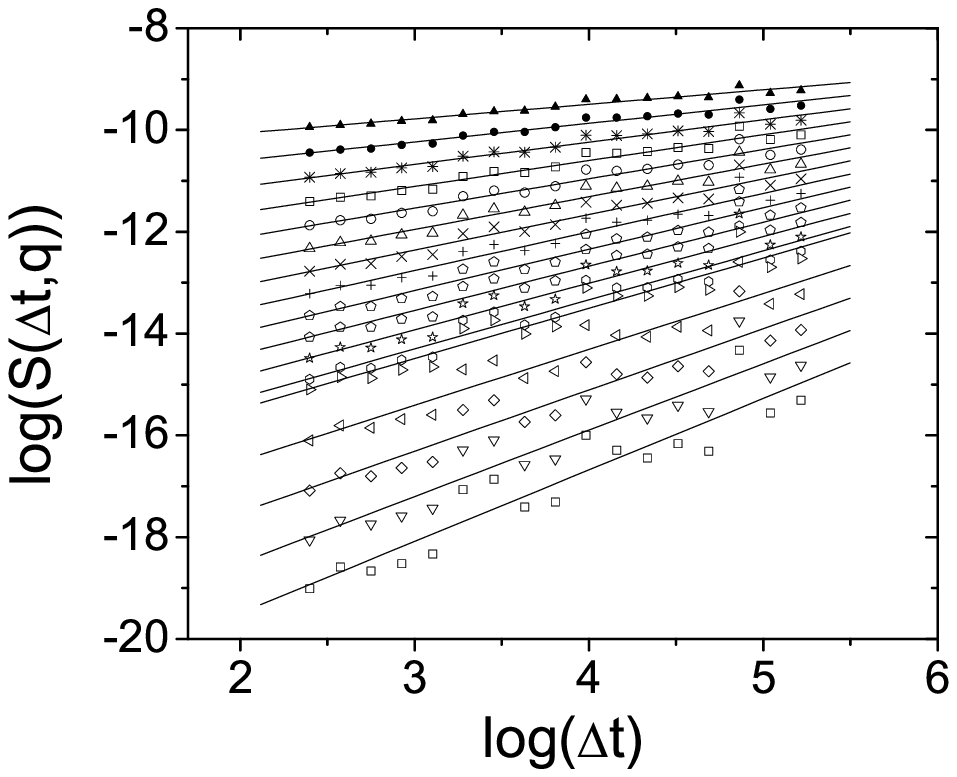,width=125pt}}}
\caption{Left: $\tau (q)$ spectrum of DAX as measured from MF-DFA(1) with the analytical curve for the fitted Lux model. Top right: Scaling of DAX for moments $q=-3,-2,-1,0.1,0.3,0.5\dots 2.5$. Bottom right: Scaling of DAX for moments $q=2.7,2.9,\dots 4.9,5,5.5,6,6.5,7.$}
\label{fig:dax-lux-tq}
\end{figure}

Let us make clear what parameters are necessary to (statistically) reproduce a given time series. Firstly, $\sigma$ is the standard deviation of log-returns. Secondly, parameter $\lambda$ decides the distribution of the $m^{(i)}_t$ multipliers, and through this the multiscaling behavior. And finally $k$ is the number of multipliers. In \cite{lux.model} two methods are proposed to determine $\lambda$, but the well-known \emph{scaling estimator} method is found to perform poorly compared with \emph{GMM estimation} (see also \cite{lux.test}). The performance of the scaling approach can be improved substantially by using the \emph{Multifractal Detrended Fluctuation Analysis} (MF-DFA) \cite{dfa.intro,dfa.kantelhardt} procedure instead of simple scaling. A low (first or second) order method already eliminates most of the instationarity in the empirical data and the resulting spectra have much less bias. Although simulated sample paths of the Lux model are stationary, even in this case MF-DFA brings a significant improvement. Results for the DAX dataset are shown in Figure \ref{fig:dax-lux-tq}, $\lambda_{DAX} =1.13206\pm0.00156$. Typical values are $\lambda = 1.01 \dots 1.30$.

It is also important to note, that $q \geq 3$ moments of $x$ do not exist in a strict sense, as a consequence of the fat tails. The empirically measured exponents originate from a finite-size effect, which is one form of apparent multifractality \cite{bouchaud.apparent}. However the inclusion of these moments does not change the fitted $\lambda$ values significantly.

The performance of a GMM estimator proposed by Lux \cite{lux.model,lux.test} is still better than MF-DFA by means of bias and root mean squared error. However the difference is much smaller, than for the traditional scaling estimator. We would like to note, that the typical bias and error in $\lambda$ that arises from MF-DFA does not change the actual behavior of the fitted moments significantly, and therefore it seems plausible, that a slight misspecification of $\lambda$ does not seriously decrease forecasting performance. Finally, the scaling method might be much more intuitive, and should not be discarded.

\begin{figure}[!t]
\hbox{\psfig{file=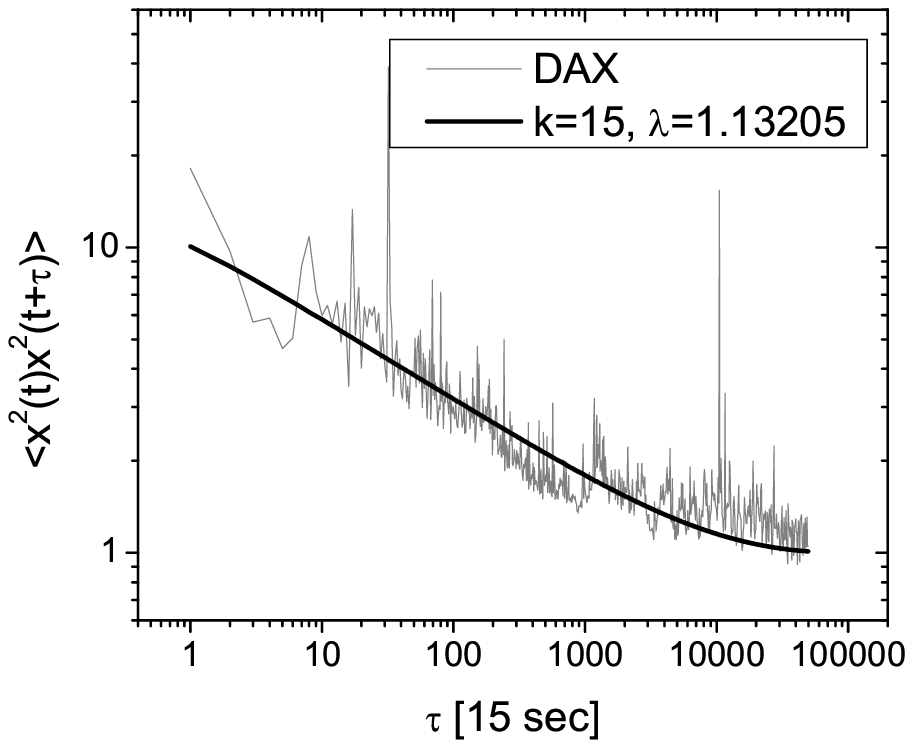,width=200pt}\psfig{file=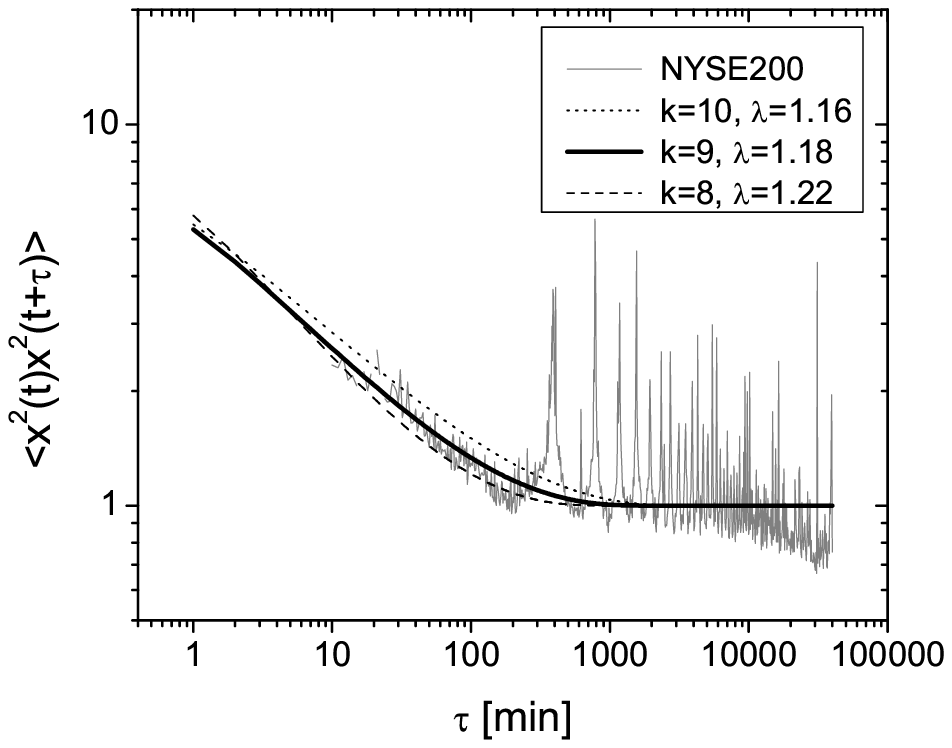,width=203pt}}
\caption{Left: Empirical $\left\langle x^2(t)x^2(t+\tau ) \right\rangle$ values for DAX and the fitted Lux model. Right: Empirical $\left\langle x^2(t)x^2(t+\tau ) \right\rangle$ values averaged over $200$ stocks in NYSE200, and possible fits from the Lux model.}
\label{fig:emp-x2x2ac}
\end{figure}

The $k$ number of multipliers is also called cascade depth. 
As $k$ basically determines the length of the power-law regime in the memory of the generated process, it is straightforward to measure $k$ from autocorrelations. Our choice is to use the second order autocorrelations of logarithmic volatility as defined in \eqref{eq:x2x2ac}. Using combinatorial assumptions Lux \cite{lux.model} gives the exact formula

\begin{eqnarray}
\left\langle x^2(t)x^2(t+\tau ) \right\rangle = \sigma^4\left ( \sum_{n=1}^{k-1}\left ( \prod_{j=1}^n \left (1 - \frac{1}{2^{k-j}}\right )^\tau \right )\left (1 - \left (1 - \frac{1}{2^{k-n-1}}\right )^\tau\right ) \right. \nonumber \\ \left. \left\langle (m_t^{(i)})^2\right\rangle^n \left\langle m_t^{(i)}\right\rangle^{2k-2n} + \left (1 - \left (1 - \frac{1}{2^{k-1}}\right )^\tau\right )\left\langle m_t^{(i)}\right\rangle^{2k} \right ),
\label{eq:analx2x2ac}
\end{eqnarray} 
where the moments can be calculated from \eqref{eq:lux-moments}. This expression can be interpreted to have two qualitatively different regimes. For $\tau < 2^k$ the correlations are power-law, and for $\tau > 2^k$ they disappear. We observed similar typical behavior in various stocks, and based on this crossover we managed to fit the $k$ cascade depth. The procedure for DAX is shown in Figure
\ref{fig:emp-x2x2ac}. We also give the average of $\left\langle x^2(t)x^2(t+\tau ) \right\rangle$ over the $200$ stocks of NYSE200. In the empirical data one can clearly see peaks corresponding to the daily periodicity, which can be filtered out by normalization of the intraday volatility distribution \cite{stanley.statistical}. This however does not affect our observation, that the fit from the Lux model is very sensitive to the value of $k$. If we first choose $\lambda$ according to the MF-DFA method proposed herein or the GMM estimator of \cite{lux.model}, then the above procedure determines the value of $k$. The appropriate choice of $k$ ensures the correct length of power-law memory in the simulated time series.

Note, that the $\Delta t$ size of the elementary time step determines the resolution of modeling. The crossover from power-law memory to uncorrelated log-returns is qualitatively at $2^k$ time steps. In a physical interpretation different resolution models of the same process (stock) should have the same length of memory, thus $2^k\Delta t = const$.
\subsection{Tail behavior}
\begin{figure}[!t]
\hbox{\psfig{file=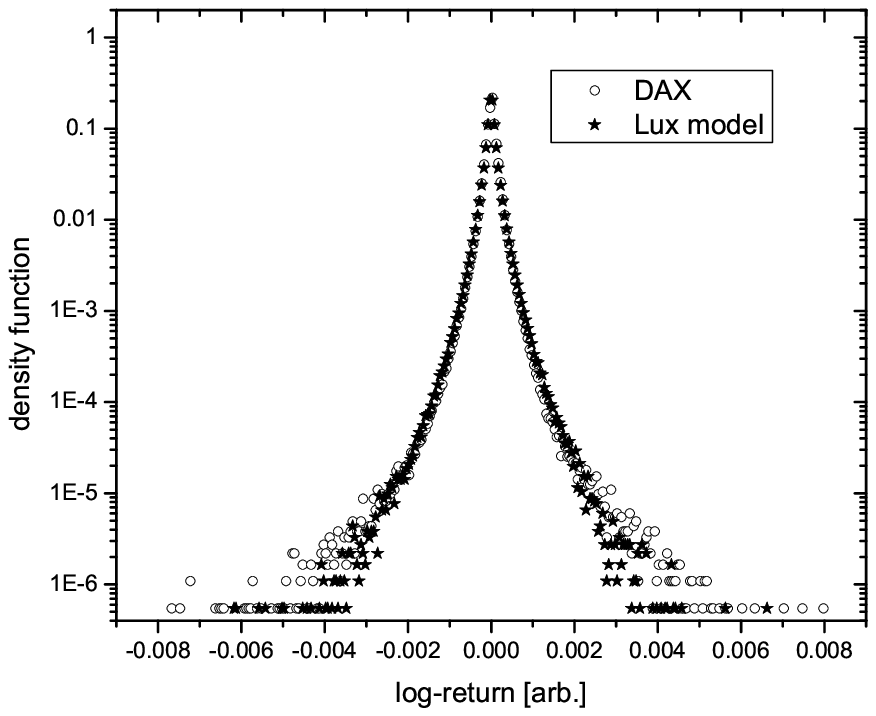,width=240pt}\vbox{\psfig{file=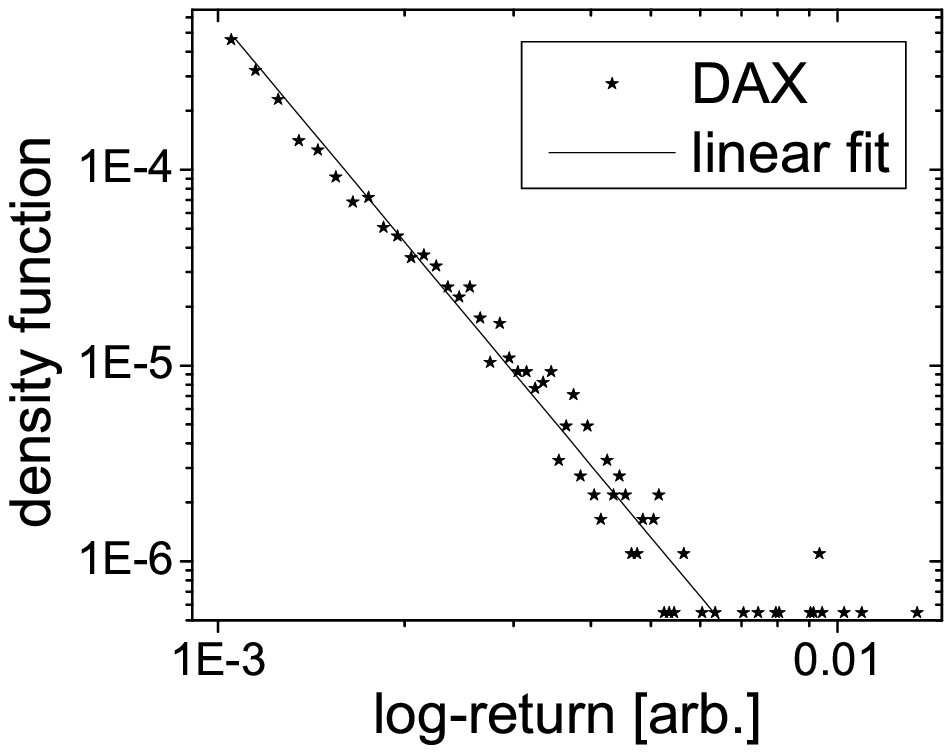,width=120pt}\\\psfig{file=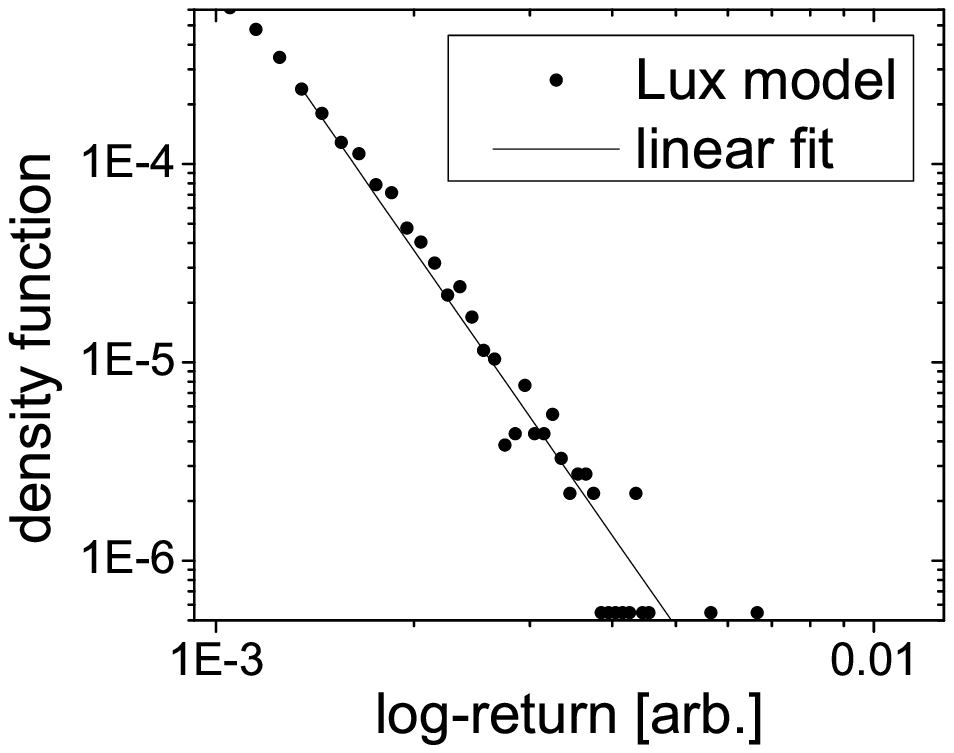,width=120pt}}}
\caption{Left: Distribution of the logarithmic returns for DAX in the period 1999-2001, and the corresponding fitted Lux model. Top right: Tail behavior of DAX logarithmic returns. The power-law fit gives a tail exponent $\alpha \approx 3.79$. Bottom right: Tail behavior of the Lux model logarithmic returns. The power-law fit gives a tail exponent $\alpha \approx 4.77$.}
\label{fig:dax-lux-dist}
\end{figure}
If we make a log-return distribution histogram (Figure \ref{fig:dax-lux-dist}) we can see that the TLF-like regime is fitted excellently by the model. Assuming tails are power-law, the tail exponent is larger than expected. For DAX $\alpha \approx 3.79$, while for the fitted Lux model it is $\alpha \approx 4.77$. This results in missing finite-sample kurtosis, $\kappa_4^{DAX} \approx 603$, $\kappa_4^{Lux}\approx 34.2$. This might limit the applicability of the model for long-time forecasting, where these many-sigma events come into play. The problem is that the value of measured kurtosis depends on the $k$ number of multipliers used. We fit $k$ from $\left\langle x^2(t)x^2(t+\tau ) \right \rangle$-type autocorrelations, the $\tau = 0$ value of which is closely related to the kurtosis. To match the $\tau > 0$ behavior of typical minute-scale data  $k\sim 10$ has to be chosen, while the correct $\tau = 0$ behavior is given by $k\sim 30$. There seems to be no solution to this problem within the framework of the Lux model, because all the parameters of the model are strictly set. It is in principle possible to introduce a new parameter or to change the distribution of the multipliers from log-normal to give superior results.

Although visual (see Figure \ref{fig:dax-lux-dist}) and statistical
methods cannot easily distinguish simulation results from empirical
power-laws in the tail of the distribution, there is good reason to
think that the simulated tail should decay faster since the random
variable results from the product of variables taken from a Gaussian
and a log-normal distribution. This is the reason why, in constrast to
the empirical distribution (``inverse cubic law''), the moments
 $\left\langle \vert x_{\Delta t}\vert^q\right\rangle$ exist for any
$q$, see \eqref{eq:lux-moments}.

\section{Third moment related effects}
\label{sec:third}
In this section we first introduce two extensions of the Lux model, for up-down asymmetry and leverage autocorrelations. Then we evaluate the implications of leverage on volatility forecasting.

The original Lux model does not reproduce the up-down asymmetry:
\begin{equation}
\left\langle x_{\Delta t}(t)^3\right\rangle \propto \left\langle{A(t)}^3 S(t)^3\right\rangle = \left\langle{A(t)}^3\right\rangle\left\langle S(t)^3\right\rangle = \left\langle{A(t)}^3\right\rangle*0=0.
\end{equation}
Similarly, leverage autocorrelations vanish:
\begin{eqnarray}
C_L(\tau ) \propto \left\langle A(t)S(t) A(t+\tau )^2S(t+\tau )^2\right\rangle = \nonumber \\
\left\langle A(t)S(t) A(t+\tau )^2S(t+\tau )^2\right\rangle \left \langle S(t)\right\rangle = 0.
\end{eqnarray}
The two main assumptions for these calculations are that $A(t)$ and $S(t)$ are independent of each other, and that $S(t)$ has a vanishing third moment. Relaxing one (or both) of these conditions can be adequate to account for these two additional stylized facts.

\subsection{The extended Lux model}

Returning to the basic idea of stochastic volatility models \eqref{eq:volat}, we can modify the sign term in the Lux model. Previously we assumed $S(t)=N(0,1)$. Since the $A(t)$ amplitude factor is non-negative, skewness can only originate from the sign. Therefore we introduce a new, discrete random variable:
\begin{equation}
\begin{array}{ll} 
{\mathbb P} \left(S=-\frac{1/\sqrt{2}}{1/2-\epsilon}\right)=1/2-\epsilon , \\
{\mathbb P} \left(S=\frac{1/\sqrt{2}}{1/2+\epsilon}\right)=1/2+\epsilon .
\end{array}
\label{eq:sign}
\end{equation}
One can show, that the formula
\begin{eqnarray}
\left\langle x^3(t)\right\rangle = \left\langle \left(A(t)\sigma S(t)\right)^3 \right\rangle \approx 
\left\langle A(t)^3 \right\rangle \sigma^3 \left\langle S(t)^3 \right\rangle = \nonumber \\ -4\epsilon \sigma^3*\exp \left (\frac{3}{4}k(\lambda -1)\ln 2\right )
\label{eq:skew.sign}
\end{eqnarray}
works well for a range of parameters and can be used to fit $\epsilon$ to empirical data. This gives a mean level of skewness in the simulations. The approach shows one typical effect that is inherent in the Lux model (and possibly other similar stochastic volatility models). Although we can reproduce the mean level of skewness (or leverage autocorrelations as we will see), the model is stationary, and consequently the predicted fluctuations are much smaller than for empirical data.

This new source of skewness does not change the time correlation properties of the process. Pochart and Bouchaud proposed a new term to introduce leverage autocorrelations to their MRW model \cite{mrw.skewness}. Their approach seems to be readily applicable to any stochastic volatility model specified as \eqref{eq:volat}. We therefore extend the Lux model as follows:
\begin{equation}
x(t) = \sqrt{2^k\prod_{i=1}^km_t^{(i)}}\exp\left(-\sum_{t'<t}S(t')K(t-t')\right) \sigma S(t).
\label{eq:skew.intro}
\end{equation}
$K(t)$ can be considered a \emph{kernel funcion}. It can be basically any kind of decaying function, but later on we will prove, that qualitatively $C_L(\tau)\sim K(\tau )$ in an appropriate limit (also see \cite{mrw.skewness}).

One can show analytically, that for 
\begin{equation}
K(t-t') = \Theta(\tau_0-(t-t'))\frac{K_0}{(t-t')^\alpha}
\label{eq:k-powerlaw}
\end{equation}
in a first order approximation the relation 
\begin{equation}
\left\langle x_{\Delta t}^2(t)\right\rangle \approx \sigma^2 \Delta t
\end{equation}
remains unchanged. Note that $\tau_0$ in \eqref{eq:k-powerlaw} is a finite time cutoff in the memory. We used it instead of an exponential factor Pochart and Bouchaud \cite{mrw.skewness} use to regularize integrals in calculations for the $\alpha < 0.5$ case, because it makes simulations run faster, while (taken large enough) not affecting the results significantly. 

In the $\alpha > 0.5$ case for the third moment we get
\begin{equation}
\left\langle x^3_{\Delta t}(t) \right\rangle \approx -\frac{6K_0\sigma^3}{2-\alpha}\exp(k\ln 2(1-\lambda)/2){\Delta t}^{2-\alpha}.
\label{eq:skew.third}
\end{equation}
Combining \eqref{eq:k-powerlaw} and \eqref{eq:skew.third} gives for skewness:
\begin{equation}
\kappa_3(\Delta t) = \frac{\left\langle x^3_{\Delta t}(t) \right\rangle}{\left\langle x^2_{\Delta t}(t) \right\rangle^{3/2}} \approx -\frac{6K_0}{2-\alpha}\exp(k\ln 2(1-\lambda)/2)\Delta t^{0.5-\alpha}.
\label{eq:skew}
\end{equation}
This shows that the addition of this new term also results in the appearance of skewness, which decays with time for $\alpha > 0.5$, and increases for $\alpha < 0.5$. Pochart and Bouchaud \cite{mrw.skewness} also report, that their process loses skewness in the continuous-time limit, when the size of the elementary time steps $\Delta t \rightarrow 0$. This is corrected by our independent skewness parameter $\epsilon$.

We do not wish to repeat all the calculations of \cite{mrw.skewness} for the case of the Lux model, only to indicate, that the explicit form of the stochastic volatility term is indifferent to the behavior of leverage autocorrelations. Instead, we focus on the applications of this approach. As defined in Section \ref{sec:stylized}, leverage autocorrelations are
\begin{equation}
C_L(\tau ) = \frac{\left\langle x(t) x^2(t+\tau )\right\rangle}{\left\langle x^2(t) \right\rangle^2}.
\end{equation}
In empirical data $C_L(\tau )$ is typically said to decay exponentially \cite{leverage1,bouchaud.leverage,leverage3}. Pochart and Bouchaud argue \cite{mrw.skewness}, that the introduction of an exponential characteristic time for leverage would destroy the scale invariance of the process, and therefore they treat the power-law leverage autocorrelations as an approximation. However, later in this section we find that the power-law also well describes data.

Now in order to see the behavior of leverage autocorrelations, we introduce a different $K(t-t')$ kernel function to \eqref{eq:skew.intro} by analogy to \cite{bouchaud.leverage}:
\begin{equation}
K(t-t')=-K_1\exp\left(-\frac{t-t'}{\tau_0}\right).
\end{equation}
We make a first-order approximation for the correlation term
\begin{equation}
\exp\left(-K_1\sum_{t'<t}S(t')\exp\left(-\frac{t'}{\tau_0}\right)\right)\approx 1-K_0\sum_{t'<t}S(t')\exp\left(-\frac{t'}{\tau_0}\right).
\label{eq:skew.linear}
\end{equation}
This holds, if the argument of the first exponential is small in absolute value. Since the series $\{S(t')\}$ is a Brownian motion, we know that qualitatively
\begin{equation}
\left\vert\sum_{t'<t}S(t')\right\vert \sim (t-t')^{1/2},
\end{equation}
We can substitute the remaining exponential with $\theta(t')\theta(\tau_0-t')$, to get an approximate condition for the weakly correlated limit:
\begin{equation}
K_0\sum_{t'<t}S(t')\exp\left(-\frac{t'}{\tau_0}\right)\approx K_0\sum_{0<\tau\leq \tau_0}S(t')\propto K_1\sqrt{\tau_0} \ll 1.
\label{eq:weak}
\end{equation}
Similar formulas can be given for power-law kernel functions defined in \eqref{eq:k-powerlaw}. 

Now we return to the notation $K(t')$ to deduce some general results. Given the above linearization is justified, leverage autocorrelations can be written as 
\begin{eqnarray}
\left\langle \sigma^2 S(t)^2\right\rangle^2 C_L(\tau ) = \left\langle x(t) x(t+\tau )^2\right\rangle = \left\langle A(t) A(t+\tau )^2 \sigma^2\right. \nonumber \\ \left.
S(t+\tau )^2 \sigma S(t)\left(1+\sum_{t'<t}S(t')K(t')\right) \left( 1+\sum_{t''<t+\tau }S(t'')K(t'')\right)^2 \right\rangle .
\label{eq:skew.levercorrel}
\end{eqnarray}
By the independence of some terms, the only contribution from the parentheses that remains (with 1, 1, and $t''=t$ terms selected from the sums respectively) is
\begin{equation}
\left\langle \sigma^2 S(t)^2\right\rangle^2 C_L(\tau ) = \left\langle A(t) A(t+\tau )^2 S(t+\Delta t)^2 \right\rangle \left\langle 2\sigma^3 S(t)^2 K(\tau ) \right\rangle ,
\end{equation}
and finally
\begin{equation}
C_L(\tau ) = \left\langle A(t) A(t+\tau)^2 \right\rangle 2 \sigma^{-1} K(\tau ) .
\end{equation}
This is a general result, that holds for any kernel function $K(t')$, as long as the linearization works. To calculate the amplitude autocorrelations, we generalize \eqref{eq:analx2x2ac} for arbitrary order autocorrelations. Straightforward calculation yields:
\begin{eqnarray}
\left\langle A(t)^\alpha A(t+\tau )^\beta \right\rangle = \sum_{n=1}^{k-1}\left ( \prod_{j=1}^n \left (1 - \frac{1}{2^{k-j}}\right )^\tau \right )\left (1 - \left (1 - \frac{1}{2^{k-n-1}}\right )^\tau\right ) \nonumber \\ \left\langle \left (m_t^{(i)}\right )^{(\alpha + \beta )/2}\right\rangle^n \left (\left\langle \left (m_t^{(i)}\right )^{\alpha /2}\right\rangle \left\langle \left (m_t^{(i)}\right )^{\beta /2}\right\rangle \right )^{k-n} + \nonumber \\  \left (1 - \left (1 - \frac{1}{2^{k-1}}\right )^\tau\right )\left ( \left\langle \left (m_t^{(i)}\right )^{\alpha /2} \right\rangle \left\langle \left (m_t^{(i)}\right )^{\beta /2} \right\rangle \right )^{k}
\label{eq:analxnxnac}
\end{eqnarray}
Skewness originating from the sign changes leverage correlations in a way, that can be calculated from \eqref{eq:skew.sign}: 
\begin{equation}
C_L(\tau ) = \left\langle A(t) A(t+\tau)^2 \right\rangle 2\sigma^{-1}K(\tau ) (1 - 2\epsilon K(\tau )).
\end{equation}
The last term is a third order correction and has no significance in practice ($\epsilon K(\tau = 1)\sim 10^{-2}$ typically). For the exponential case this expression performs similar to those presented in the recent comparison by Perello and Masoliver \cite{leverage4}.

\begin{figure}[!t]
\hbox{\psfig{file=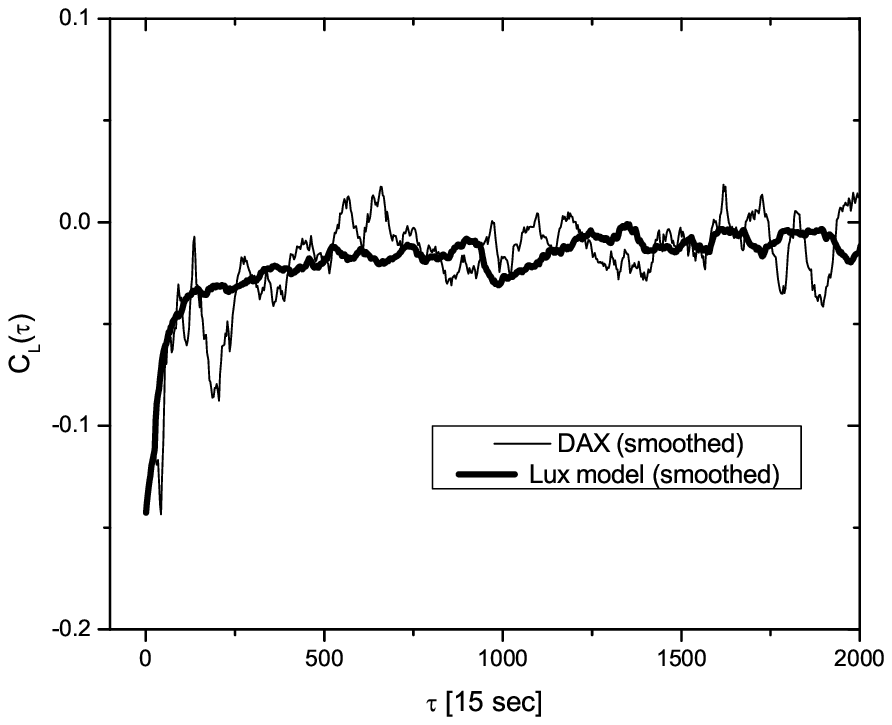,width=250pt}\vbox{\psfig{file=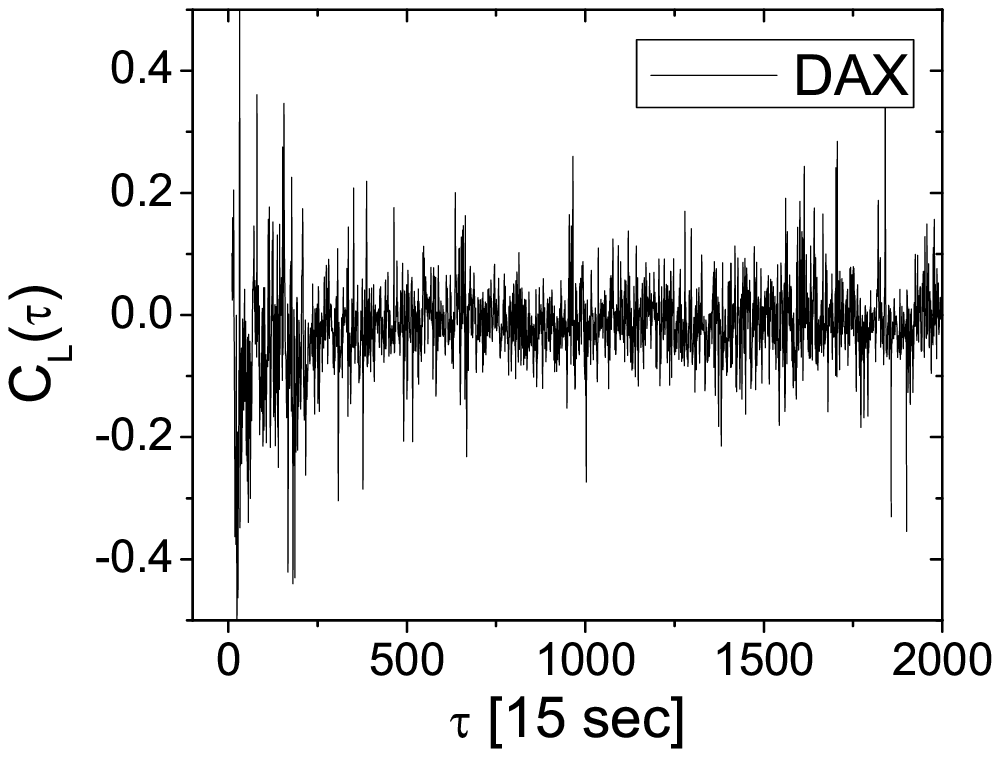,width=125pt}\\\psfig{file=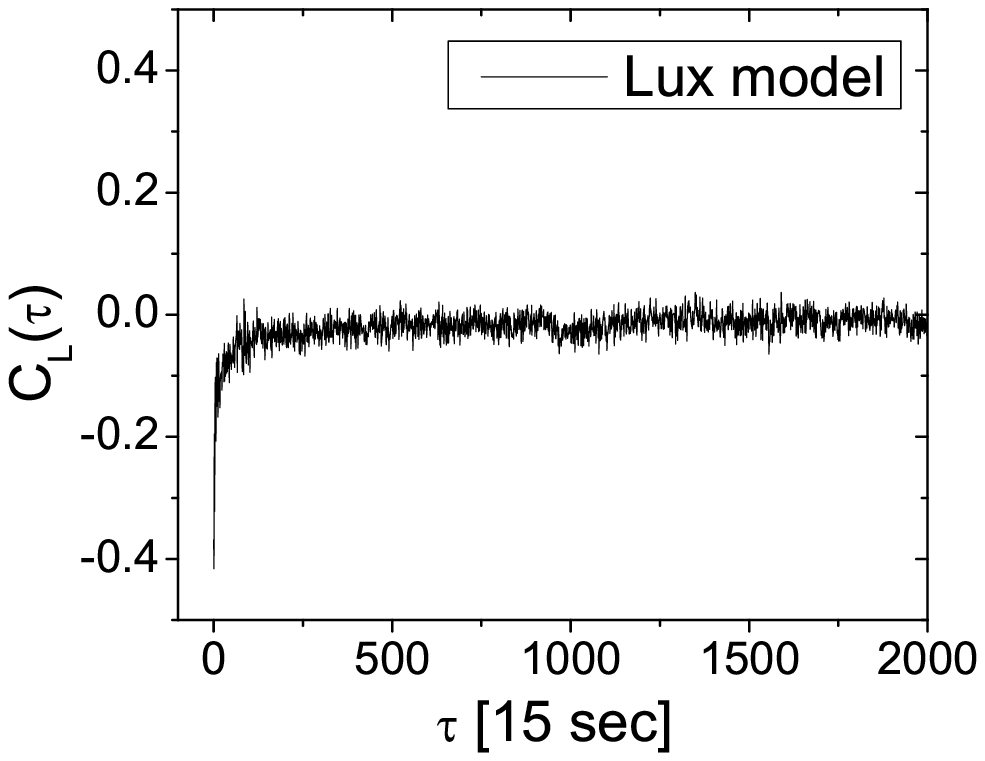,width=125pt}}}
\caption{Left: Measured leverage autocorrelations of DAX and a sample path with fitted parameters. Noise was removed by a $50$-neighbor averaging. Top right: Leverage autocorrelations of DAX without smoothing. Bottom right: Leverage autocorrelations of simulated sample path fitted to DAX, without smoothing.}
\label{fig:dax-lux-lev}
\end{figure}

We could find no indisputable evidence for a well-defined, mean leverage in NYSE200, only for individual stocks.  We found DAX leverage autocorrelations to be well fitted by a power-law type kernel, with $\alpha \approx 0.384$, $K_0 \approx 0.03$. Results can be seen in Figure \ref{fig:dax-lux-lev}. From the comparison of the smoothed graphs one can see, that the mean behavior of the model and the original time series is very similar. However if we do not remove the noise by smoothing, it can be seen, that the empirical fluctuations around the mean value are much larger than the Lux model would predict. Part of this effect can originate from that the data was unfiltered. The predictable intra-daily trends \cite{stanley.statistical} of volatility that are not reproduced by the model can account for increased fluctuations. Further investigation is possible, here we would only like to note on the fact, that there is no a priori reason why the Lux model should give the right size of fluctuations. Possibly a higher inherent level of noise is present in more complicated, agent-based simulations. In our opinion empirically observed fluctuations are larger, because of the inevitable instationarity of the price process. This effect seems to be similar to that reported in \cite{rally}. Third-moment related quantities are instationary, and this might limit their applicability to forecasting.

\subsection{The role of leverage in volatility forecasting}

At present maximum likelihood type estimators for our volatility forecasts are unavailable. As noted in Section \ref{sec:causality}, a partial success has been achieved in \cite{cf.regime}, but the results are not readily applicable to complicated cases like the Lux model or MRW. Lux \cite{lux.model} proposes the use of a Levinson-Durbin algorithm (algorithm 6 of \cite{levdurb}) instead of maximum likelihood. This can be proven to be the best linear predictor, and it takes power-law autocorrelations of logarithmic returns into consideration. A simple formalization, consistently with our notations would be: 
\begin{equation}
\tilde x^2(t+h) = \sum_{i=0}^{m-1}\phi_{i,m}^{(h)}\tilde x^2(t-i),
\label{eq:levdurb}
\end{equation}
where $\tilde x^2 = x^2-\sigma^2$, $\sigma$ is the standard deviation of log-returns and $\phi_{i,m}^{(h)}$ are constants determined by the set $\tilde x^2(t-m+1\dots t)$. For explicit formulas see \cite{levdurb}.

\begin{figure}[!tb]
\centerline{\psfig{file=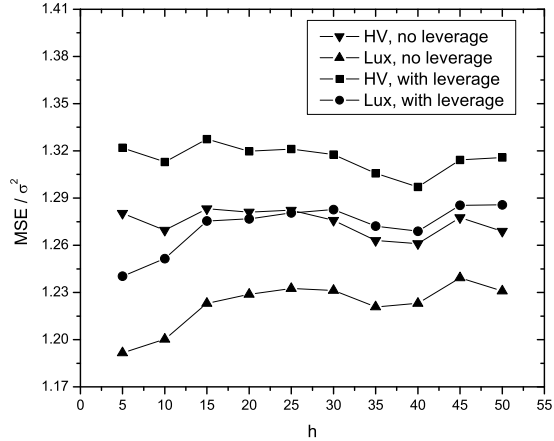,height=200pt}}
\caption{Comparison of the mean sqared error of volatility forecasts by historical volatility and Levinson-Durbin ($m=2000$), when the underlying time series is generated by a Lux model including and not including leverage. The approximation is given with $30000$ sample runs, model parameters are $k=12$, $\lambda = 1.1$, $K_0 = 0.03$, $\alpha = 0.384$. The results shown are from a $5$-neighbor smoothing of a $h=1,2,\dots,50$ series.}
\label{fig:lux-eval-lever}
\end{figure}

This method ignores the sign process. Now we will show, that the sign information and in particular the leverage effect are indeed relevant to forecasting performance. We carried out measurements to compare the (theoretical) mean squared errors of different methods in the presence of leverage autocorrelations. We did this by running $30000$ Monte Carlo iterations with sample paths simulated by the Lux model. The parameters were fitted to DAX ($\lambda = 1.13205$, $k=15$), but we also used the respective values of leverage ($K_0 = 0.03$, $\alpha = 0.384$). The results (Figure \ref{fig:lux-eval-lever}) show that the effect of asymmetry is not negligible from the point of view of volatility forecasting, because the accuracy of the forecasts is decreased. Actually the loss of accuracy caused by leverage is roughly the same as the gain from historical volatility to the Lux model. This shows, that leverage autocorrelations do not only produce asymmetry in volatility smiles, but are also relevant for the predictions of future conditional volatility. Note that this loss also applies to historical volatility, so it does not affect the relative performance of the two methods.

\begin{figure}[!t]
\centerline{\hbox{\psfig{file=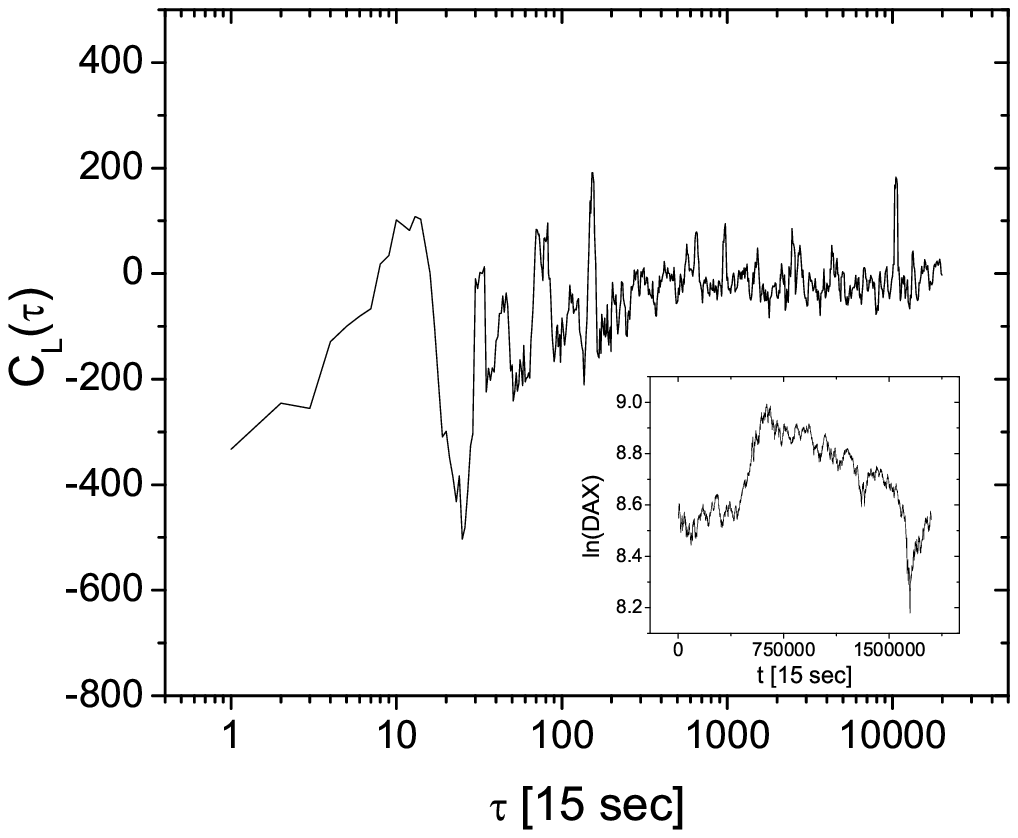,width=230pt}\psfig{file=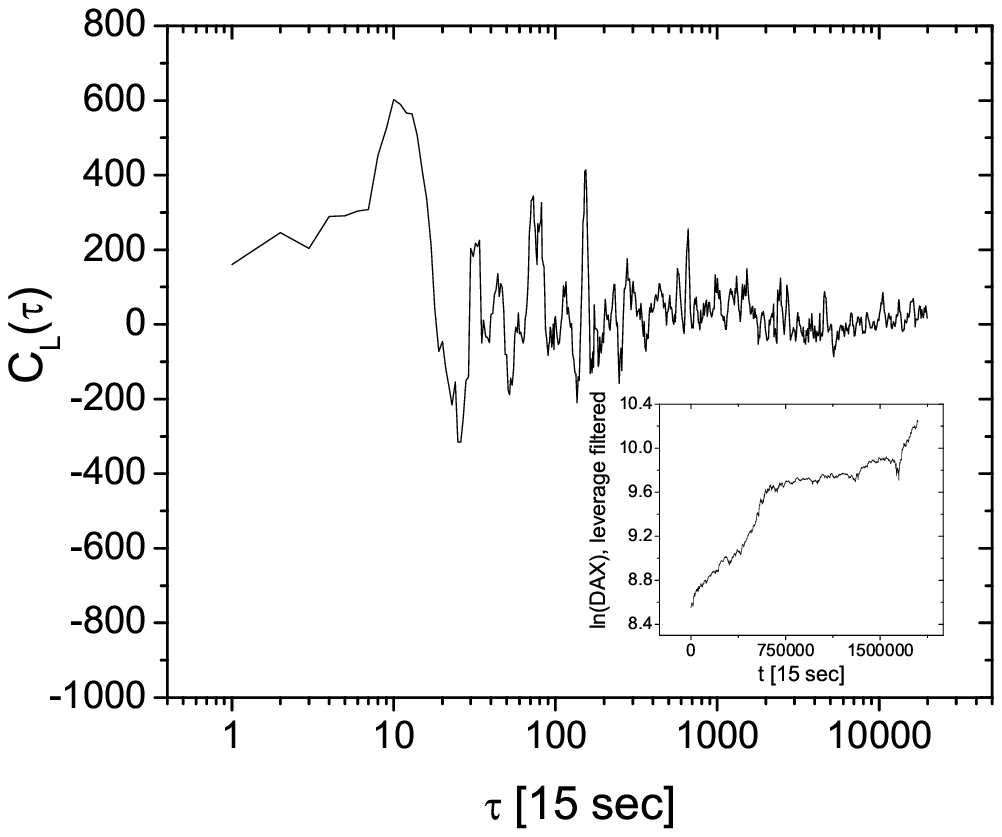,width=230pt}}}
\caption{Left: Leverage of natural logarithm of DAX. Right: Leverage of natural logarithm of DAX after the application of the leverage filter. $K(t-t')=\frac{0.03}{(t-t')^{0.384}}$. The insets show the time series.}
\label{fig:dax-delev}
\end{figure}

\begin{figure}[!t]
\centerline{\psfig{file=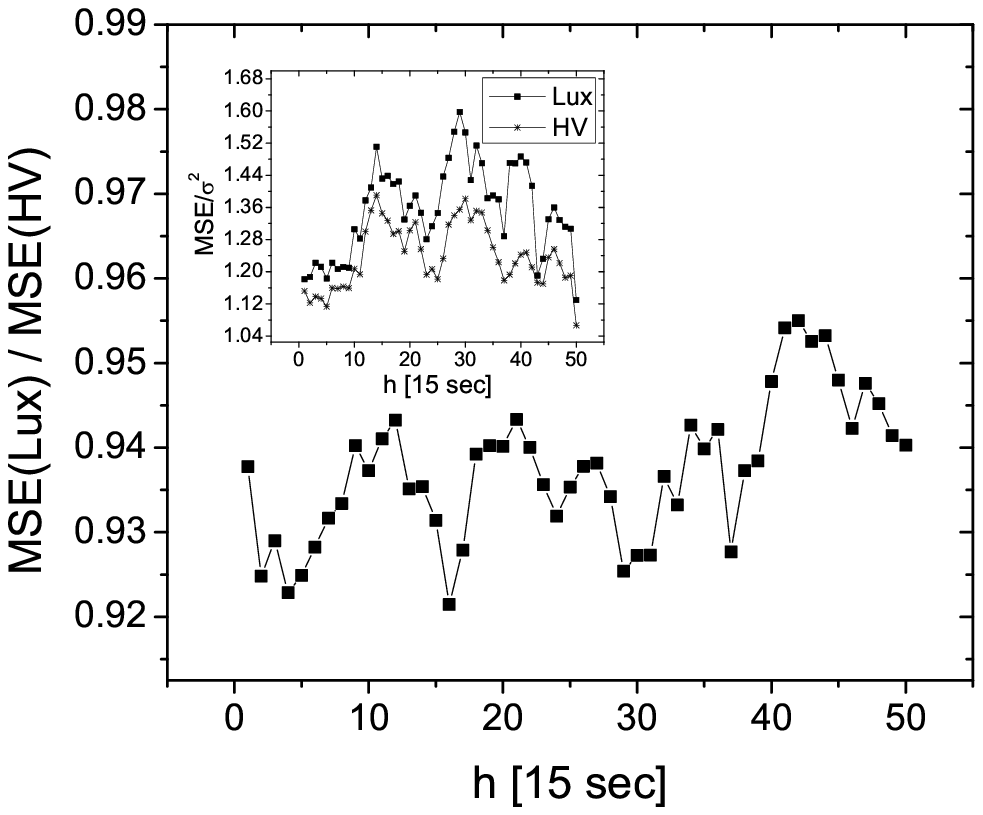,width=200pt}\psfig{file=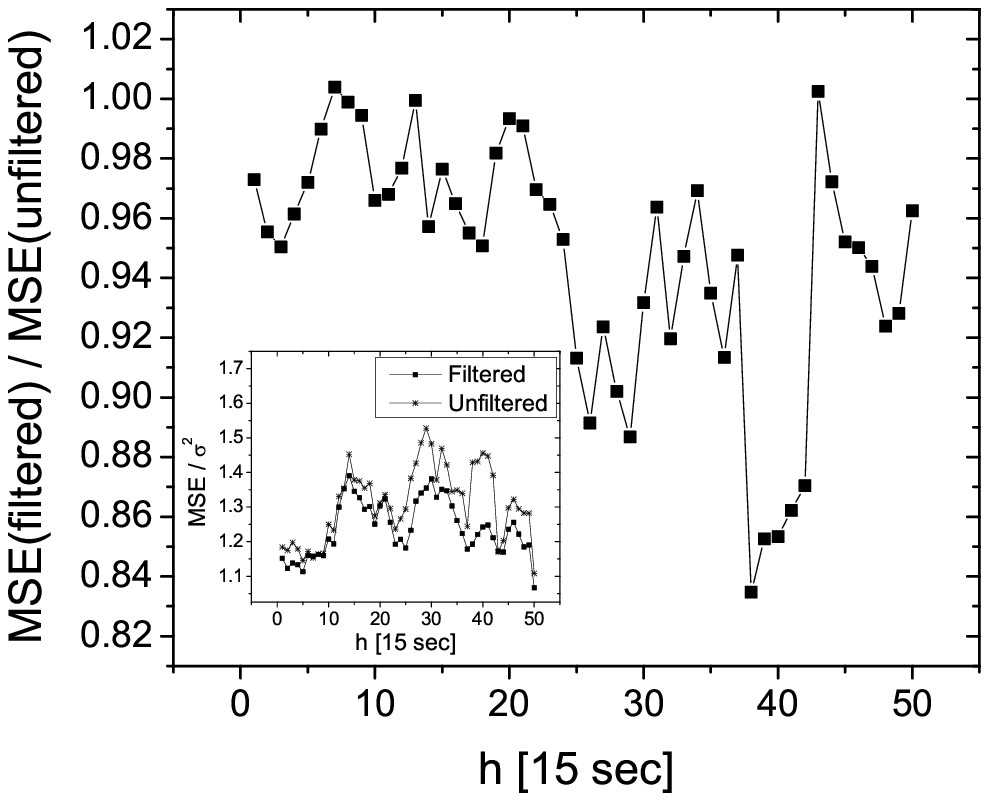,width=200pt}}
\caption{Left: performance of the linear forecast versus historical volatility for the leverage filtered DAX series. The inset shows the mean squared error values with a 5-neighbor smoothing. Right: comparison of linear forecasts for the filtered and the unfiltered data. The inset shows the mean squared error values with a 5-neighbor smoothing. All results are normalized with the respective standard deviations.}
\label{fig:lux-eval-delev}
\end{figure}

Let us now consider the empirical time series to originate from any stochastic volatility model \eqref{eq:volat}, where $S(t)$ is a discrete, symmetric ($\epsilon = 0$) sign process. It is straightforward to construct a simple filter to remove the leverage effect from the logarithmic returns:
\begin{equation}
\tilde x(t) = x(t)\exp\left ( \sum_{t'<t}S(t')K(t-t')\right),
\label{eq:delev}
\end{equation}
where $x(t)$ is the original set of log-returns, $K(t-t')$ is the leverage kernel function fitted to the data, and $S(t')=sgn\left( x(t')\right)$. Figure \ref{fig:dax-delev} shows the leverage of the natural logarithm of DAX and its filtered counterpart, with the insets showing the actual time series. After the application of the filter, the resulting data are up-down symmetric, except for the drift of prices. This property improves the performance of both historical volatility and Lux model-based forecasts, which originally both omit sign information. The results are given in Figure \ref{fig:lux-eval-delev}. The comparison between forecasts on the filtered and unfiltered data shows that there is a reasonable gain in accuracy. The average decrease of error\footnote{These results are related to the performance of a procedure, that first filters out past leverage autocorrelations, then makes a forecast based on this data, and then restores leverage autocorrelations. The only difference is that to restore future leverage, we would have to use "future" sign information. This can be formally resolved by taking $S(t' > t) = 0$. Further research will be necessary to evaluate this in detail.} is comparable to the improvement from historical volatility to the Lux model. This underlines the theoretical expectations (Figure \ref{fig:lux-eval-lever}).

\section{Conclusion}
In this paper we have reviewed the generalized stochastic volatility
framework used to introduce multifractal models of stock prices. We
focused our investigations on the Lux model \cite{lux.model} and we have shown one efficient procedure to fit the model parameters to actual data. We have given arguments, that the effective length of the power-law memory (one stylized fact that can be measured in practice) in the model is governed by the parameter $k$. We have introduced an extension of the Lux model to give up-down asymmetry, and also adopted the approach of Pochart and Bouchaud \cite{mrw.skewness} to account for leverage autocorrelations. This extended model reproduces the mean behavior of real stock market data well, but not the large fluctuations resulting from instationarity. We have introduced a filter to remove this leverage from the data. Filtered DAX data and Monte Carlo simulations both show, that the presence of asymmetry increases mean squared error of forecasts that omit sign information. Consequently, it can be useful to extend the models in use, taking leverage into consideration with the hope of substantial gain in accuracy. Further research should aim to produce new techniques, that use the available sign information efficiently to take advantage of our findings.


\begin{thebibliography}{00}
\bibitem{bachelier} L. Bachelier: Théorie de la speculation, Annales Scientifiques de l'Ecole Normale Supérieure {\bf III-17}, pp. 21-86 (1900)
\bibitem{mandelbrot.certain} B. Mandelbrot: The Variation of Certain Speculative Prices, Journal of Business 36, pp. 307-332 (1963)
\bibitem{stanley.universality2}
H. E. Stanley and Vasiliki Plerou: Scaling and universality in economics: empirical results and theoretical interpretation, Quantitative Finance 1, pp. 563-567 (2001)
\bibitem{stanley.contribute}
H.E. Stanley, L.A.N. Amaral, D. Canning, P. Gopikrishnan, Y. Lee, Y. Liu: Econophysics: Can physicists contribute to the science of economics?, Physica {\bf A} 269, pp. 156-169 (1999)
\bibitem{ivory}
J. Doyne Farmer: Physicists attempt to scale the ivory towers of finance, International Journal of Theoretical and Applied Finance 3, pp. 311-333 (2000)
\bibitem{vicsek.turb} T. Vicsek, A.-L. Barabási: Multi-affine model for the velocity distribution in fully
turbulent flows, J. Phys. A: Math. Gen. {\bf 24}, pp. 845-851 (1991)
\bibitem{mandelbrot.mmar} B. Mandelbrot, A. Fisher, L. Calvet: A Multifractal Model of Asset Returns, Cowles Foundation
Discussion Paper 1164 (1997)
\bibitem{lux.model} T. Lux, The Multi-Fractal Model of Asset Returns: Its Estimation via GMM and Its Use for
Volatility Forecasting, University of Kiel, Working Paper (2003)
\bibitem{mrw.pre}
E. Bacry, J. Delour and J. F. Muzy: Multifractal random walk, Phys. Rev. E {\bf 64}, 26103 (2001)
\bibitem{schmitt}
Francois Schmitt, Daniel Schertzer and Shaun Lovejoy: Multifractal analysis of foreign exchange data, Applied Stochastic Models and Data Analysis 15, pp. 29-53 (1999)
\bibitem{arch} Robert F. Engle: Autoregressive Conditional Heteroskedasticity with Estimates of the Variance of United Kingdom Inflation, Econometrica 50, pp. 987-1008 (1982)
\bibitem{garch} T. Bollerslev: Generalized Autoregressive Conditional Heteroskedasticity, J. Econometrics {\bf 31}, pp. 307-321 (1986)
\bibitem{figarch} Richard T. Baillie, Tim Bollerslev, Hans Ole Mikkelsen: Fractionally integrated generalized autoregressive conditional heteroskedasticity, Journal of Econometrics 74, pp. 3-30 (1996)
\bibitem{vicsek.book} T. Vicsek, Fractal Growth Phenomena, World Scientific Publishing (1992)
\bibitem{dfa.intro} C.-K. Peng, S. V. Buldyrev, S. Havlin, M. Simons, H. E. Stanley, A. L. Goldberger: Mosaic organization of DNA nucleotides, Phys. Rev. E {\bf 49}, pp. 1685-1689 (1994)
\bibitem{dfa.kantelhardt} Jan W. Kantelhardt, Stephan A. Zschiegner, Eva Koscielny-Bunde, Shlomo Havlin, Armin Bunde, H. Eugene Stanley: Multifractal detrended fluctuation analysis of nonstationary time series, Physica A 316, pp. 87-114 (2002)
\bibitem{cf.forecasting}
Laurent Calvet, Adlai Fisher: Forecasting multifractal volatility,
Journal of Econometrics 105, pp. 27-58 (2001)
\bibitem{stanley.trading2} Vasiliki Plerou, Parameswaran Gopikrishnan, Xavier Gabaix, Luís A Nunes Amaral and H. Eugene Stanley: Price fluctuations, market activity and trading volume, Quantitative Finance 1, pp. 262-269 (2001)
\bibitem{leverage1}
Jean-Philippe Bouchaud, Marc Potters: More stylized facts of financial markets: leverage effect and downside correlations, Physica A 299, pp. 60-70 (2001)
\bibitem{mrw.skewness}
Benoit Pochart, Jean-Philippe Bouchaud: The skewed multifractal random walk with applications to option smiles, arXiv:cond-mat/0204047 (2002)
\bibitem{leverage3}
Jaume Masoliver, Josep Perello: A correlated stochastic volatility model measuring leverage and other stylized facts,
Int. J. Th. App. Fin. 5, pp. 541-562 (2002)
\bibitem{leverage4}
Josep Perello, Jaume Masoliver, Napoleon Anento: A comparison between several correlated stochastic volatility models,
arXiv:cond-mat/0312121 (2004)
\bibitem{taq}
The Trades and Quotes Database for 2000-2002, New York Stock Exchange, New York (2003)
\bibitem{stanley.book} R. N. Mantegna and H. E. Stanley: An Introduction to Econophysics: Correlations and Complexity in Finance (Cambridge University Press, Cambridge 2000)
\bibitem{bouchaud.book} J. P. Bouchaud and M. Potters: Theory of Financial Risk (Cambridge University Press,
Cambridge 2000)
\bibitem{kertesz.book} J. Kertész, I. Kondor (Eds.), Econophysics: An Emerging Science, {\emph http://newton.phy.bme.hu/$\sim$kullmann/Egyetem/konyv.html} (1998)
\bibitem{cont.stylized}
Rama Cont: Empirical properties of asset returns: stylized facts and statistical issues, Quantitative Finance 1, pp. 223-236 (2001)
\bibitem{stanley.statistical}
Yanhui Liu, Parameswaran Gopikrishnan, Pierre Cizeau, Martin Meyer, Chung-Kang Peng and H. Eugene Stanley: Statistical properties of the volatility of price fluctuations, Phys. Rev. E 60, 1390 (1999)
\bibitem{stanley.scaling}
Parameswaran Gopikrishnan, Vasiliki Plerou, Luis A. Nunes Amaral, Martin Meyer, and H. Eugene Stanley: Scaling of the distribution of fluctuations of financial market indices, Phys. Rev. E 60, 5305 (1999)
\bibitem{stanley.individual}
Vasiliki Plerou, Parameswaran Gopikrishnan, Luis A. Nunes Amaral, Martin Meyer and H. Eugene Stanley: Scaling of the distribution of price fluctuations of individual companies
Phys. Rev. E 60, 6519 (1999)
\bibitem{kertesz.drops}
A.G. Zawadowski, R. Karadi, J. Kertesz: Price drops, fluctuations, and correlation in a multi-agent model of stock markets, Physica A 316, pp. 403-413 (2002)
\bibitem{rally}
F. Lillo and R.N. Mantegna: Symmetry alteration of ensemble return distribution in crash and rally days of financial markets, Eur. Phy. J. B 15, pp. 603-606 (2000)
\bibitem{kullmann1}
L. Kullmann, J. Töyli, J. Kertész, A. Kanto, K. Kaski: Characteristic times in stock market indices, Physica A 269, pp. 98-110 (1999)
\bibitem{kullmann2}
L. Kullmann, J. Töyli, J. Kertész, A. Kanto, K. Kaski: Break-down of scaling and convergence to Gaussian distribution in stock market data, International Journal of Theoretical and Applied Finance 3, pp. 371-373 (2000)
\bibitem{stanley.commodities}
K. Matia, Y. Ashkenazy, H. E. Stanley: Multifractal properties of price fluctuations of stocks and commodities,
Europhys. Lett. 61, pp. 422-428 (2003)
\bibitem{mandelbrot.exchange} B. Mandelbrot, A. Fisher, L. Calvet: Multifractality of Deutschemark / US Dollar Exchange Rates, Cowles Foundation Discussion Paper 1165 (1997)
\bibitem{ausloos}
M. Ausloos: Statistical physics in foreign exchange currency and stock markets, Physica A 285, pp. 48-65 (2000)
\bibitem{bouchaud.leverage}
Jean-Philippe Bouchaud, Andrew Matacz and Marc Potters: Leverage Effect in Financial Markets: The Retarded Volatility Model, Phys. Rev. Lett. 87, 228701 (2001)
\bibitem{mandelbrot.large} B. Mandelbrot, A. Fisher, L. Calvet: Large Deviations and the Distribution of Price Changes, Cowles Foundation Discussion Paper 1166 (1997)
\bibitem{mrw.epjb}
J.F. Muzy, J. Delour, and E. Bacry: Modelling fluctuations of financial time series: from cascade process to stochastic volatility model, Eur. Phys. J. B {\bf 17}, pp. 537-548 (2000)
\bibitem{mrw.physa}
E.Bacry, J.Delour, J.F.Muzy: Modelling financial time series using multifractal random walks, Physica A {\bf 299}, pp. 84-92 (2001)
\bibitem{endoexo}
D. Sornette, Y. Malevergne, J.-F. Muzy: Volatility Fingerprints of Large Shocks: Endogeneous Versus Exogeneous, arXiv:cond-mat/0204626 (2003)
\bibitem{timedef1}
Eric Ghysels, Joanna Jasiak: Stochastic Volatility and Time Deformation: An Application to Trading Volume and Leverage Effects, Preprint CRDE No 95s-31 (1995)
\bibitem{timedef2}
Eric Ghysels, Christian Gouriéroux, Joanna Jasiak: Trading Patterns, Time Deformation and Stochastic Volatility in Foreign Exchange Markets, Preprint CRDE No 95s-42 (1995)
\bibitem{mantegna.trading} Giovanni Bonnano, Fabrizio Lillo, Rosario N. Mantegna: Dynamics of the number of trades of financial securities: Physica A 280, pp. 136-141 (2000)
\bibitem{cf.regime}
Laurent Calvet, Adlai Fisher: Regime-Switching and the Estimation of Multifractal Processes, working paper (2003)
\bibitem{lux.test} T. Lux, Multi-Scaling Properties of Asset Returns: An Assessment of the Power of the 'Scaling Estimator', University of Kiel, Working Paper (2003)
\bibitem{bouchaud.apparent}
J.-P. Bouchaud, M. Potters, M. Meyer: Apparent multifractality in financial time series,
Eur. Phys. J. B 13, pp. 595-599 (2000)
\bibitem{levdurb}
P.J. Brockwell, R. Dahlhaus: Generalized Levinson-Durbin and Burg algorithms
Journal of Econometrics 118, pp. 129-149 (2004)
\end{thebibliography}
\end{document}